# Universal focal reducer for small telescopes

V.L.Afanasiev[1] | V.R.Amirkhanyan[1,2] | R.I.Uklein*[1] | A.E.Perepelitsyn[1] | E.A.Malygin[1] | E.S.Shablovinskaya[1] | I.V.Afanasieva[3]

[1] Laboratory of Spectroscopy and Photometry of Extragalactic Objects, Special Astrophysical Observatory RAS, Karachai-Cherkessia, Russia
[2] Department of Radioastronomy, Steinberg Astronomical Institute of Moscow State University, Moscow, Russia
[3] Advanced Design Laboratory, Special Astrophysical Observatory RAS, Karachai-Cherkessia, Russia

**Correspondence**
*Roman Uklein. Email: uklein.r@gmail.com

**Funding Information**
Russian Scientific Foundation, 2012-00030. Ministry of Science and Higher Education of the Russian Federation, 05.619.21.0016.

This paper is devoted to the memory of Dr. Victor Afanasiev and his immense legacy. The report highlights the capabilities of two new instruments tested at the 1-meter Zeiss-1000 telescope of SAO RAS: the Stokes Polarimeter (StoP) and the MAGIC focal reducer. Optimized for the study of active galactic nuclei (AGN), methodically, these instruments are suitable for a wide range of small telescope tasks. The fields of view of StoP and MAGIC are 6' and 13' for direct images, respectively. The StoP device allows one to conduct photometric observations and polarimetric ones with a double Wollaston prism; the spectral mode was added to MAGIC. For a starlike target up to 14 mag in medium-band filters with a seeing of 1 for 20 minutes of total exposure, the photometry accuracy is better than 0.01 mag and the polarization accuracy is better than 0.6%. The available spectral range obtained with the volume phase holographic grating in MAGIC is 4000-7200AA with a dispersion of 2A/px. StoP and MAGIC received the first light in 2020 and are used in test mode at the Zeiss-1000. The report discusses the first results obtained by the authors with new instruments, as well as further prospects.

## 1   INTRODUCTION

Laboratory of Spectroscopy and Photometry of Extragalactic Objects (LSPEO) of Special Astrophysical Observatory has extensive experience in astronomical instrumentation. The leading role in the design and development of the devices was played by Dr. Victor Leonidovich Afanasiev. Mostly focused on the observations with the 6-m telescope, we have always stayed interested in small telescopes. In this report, we focus only on some of the equipment designed for these purposes.

Looking through the history of the instruments constructed ever in the laboratory, the names of the devices were often figurative and even funny. TAZIK-1[1] (Amirkhanyan, Vikul'ev, Vlasyuk, & Stepanian, 2000) is an automatic photometer that has become the forerunner of remote and automated photometric observations with optical telescopes in SAO. ADAM (Afanasiev, Dodonov, Amirkhanyan, & Moiseev, 2016) is low-and medium-resolution spectrograph for 1.6-m Russian AZT-33IK telescope. The Mapper of Narrow Galaxy Lines, or MaNGaL[2] (Moiseev, Perepelitsyn, & Oparin, 2020) was developed and manufactured for the 1-m telescope of SAO and the 2.5-m telescope of the Sternberg Astronomical Institute of the Moscow State University. All these instruments were created for specific tasks and now are efficiently working.

For the study of active galactic nuclei, an important role is played by polarimetric instruments. To measure linear polarization with an accuracy of about 0.1%, it is necessary to use simultaneous measurement of 3 Stokes parameters. A pilot project of an instrumental solution was developed by us in 2012. However, the project was suspended for several reasons. In 2019, the work continued, the upgraded device was named *Stokes Polarimeter*, or StoP. Since 2020, MAGIC was designed in the frame of *Monitoring of Active Galaxies by Investigation of their Cores* project. In this report, we briefly describe StoP as a working prototype of polarimetric scheme and universal focal reducer MAGIC itself, its observational features and perspectives.

---

[1] 'tazik' means a basin in Russian

[2] 'mangal' means a Caucassian and Middle-East barbeque.

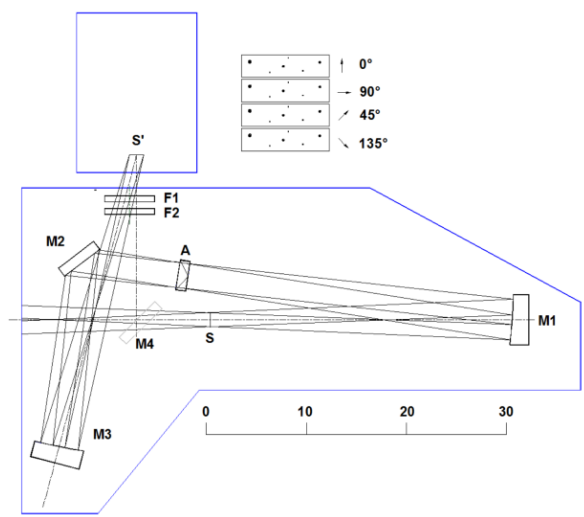

**FIGURE 1** Scheme of the Stokes Polarimeter. The scale at the bottom of the figure is given in centimeters.

## 2 STOKES POLARIMETER

The *Stokes Polarimeter* was developed as a photometer and polarimeter for the 1-meter telescope (Afanasiev, Shablovin-skaya, Uklein, & Malygin, 2021). The characteristics of the Zeiss-1000 telescope and its other instruments are described by Komarov et al. (2020). The main task of the StoP instrument is high-precision polarimetric observations of star-like targets with meter-class telescopes. For the photometric mode, two filter turrets are used, and for the polarimetry, a double Wollaston prism is introduced into the beam. The StoP device is relatively lightweight. However, due to the lateral mounting of the CCD, additional counterweights are required, which totally resulted in a total mass of 54 kg along with the internal rotation table.

Fig. 1 shows the optical layout of StoP, which corresponds to the Czerny-Turner scheme in the polarimetric mode: M1, M3 — off-axis parabolic mirrors; M2, M4 — diagonal flat mirrors; A — polarization analyzer; F1 and F2 are turrets with replaceable filters. The blue solid line shows the dimensions of the device and the light receiver. A double Wollaston prism is used as a polarization analyzer (see Oliva, 1997). The first prism is made in such a way that an ordinary and the extraordinary rays correspond to the directions of oscillations of the electric vector of $0°$ and $90°$, while for the second prism these directions correspond to $45°$ and $135°$. Double Wollaston prisms for StoP and MAGIC are made within different designs and separate the polarized images in different ways (see Fig. 3 ).

Operation in the photometer mode is provided by introducing a diagonal mirror M4 into the beam (see Fig. 1 ). There are two filter turrets F1 and F2, each with nine positions. The diameter of the filter is 50 mm. We use broadband filters and medium-band interference filters (SED)[3] .

For more details, see the main instrument article by Afanasiev et al. (2021).

## 3 MAGIC FOCAL REDUCER

### 3.1 Scheme

The StoP observations and methodical developments have shown pros and some cons, and therefore in 2020 a new scheme was designed, which became MAGIC.

The MAGIC focal reducer realizes three observational modes: photometry, long-slit spectroscopy and image polarimetry with a quadruple Wollaston prism. MAGIC weighs 23 kg and does not need counterweights due to its relatively symmetrical system, plus the CCD weighs 7 kg. MAGIC has two filter turrets with 9 positions in each, and also a special carriage for changing modes, which contains an empty hole, a VPHG grism for spectroscopy and a quadruple Wollaston prism.

In the MAGIC scheme (Fig. 2), the light from the telescope enters from the top and passes through the filter turrets. Next, there is the collimator with the focusing mechanism. In the heart of MAGIC is the mode changing carriage with the VPHG grism and the Wollaston prism. After the mode carriage light comes through the camera to the CCD detector. The possibility of three observing carriage modes makes MAGIC a versatile tool for many tasks. In addition, the empty carriage position can be used for a new element in the parallel beam to expand the MAGIC potential.

Idea of a focal reducer was developed by George Courtes (see e.g. Courtes, 1964). Concerning the modern CCD detectors, which have a limited size, the image in the focus of the telescope is usually not optimized, and needs spatial reduction to increase the field of view and decrease oversampling. The instruments of the SCORPIO family are examples of universal focal reducers on the 6-meter telescope (Afanasiev & Moiseev, 2005, 2011). In this case, the telescope focal ratio is reduced from the original $F/4$ to $F/2.6$. For observations in the dark and grey nights with the 6-m telescope, the SCORPIO instruments still play a key role.

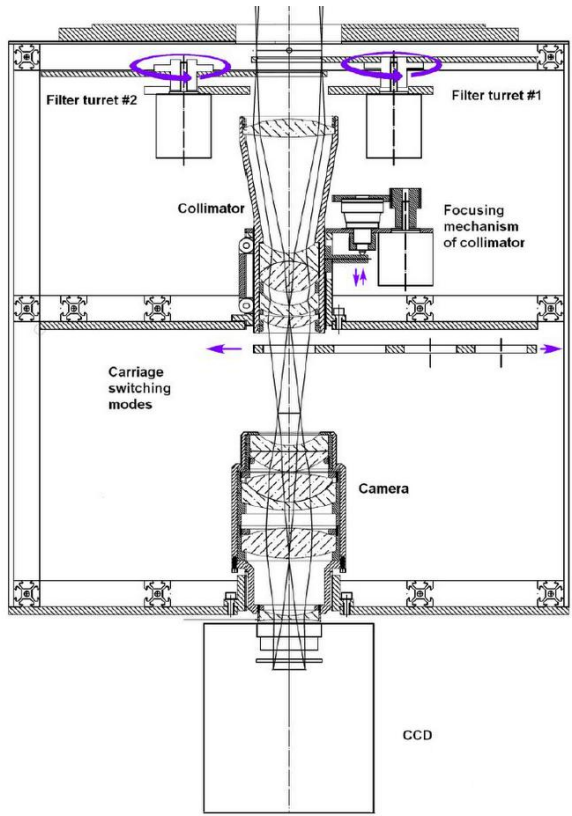

**FIGURE 2** Scheme of the MAGIC universal focal reducer.

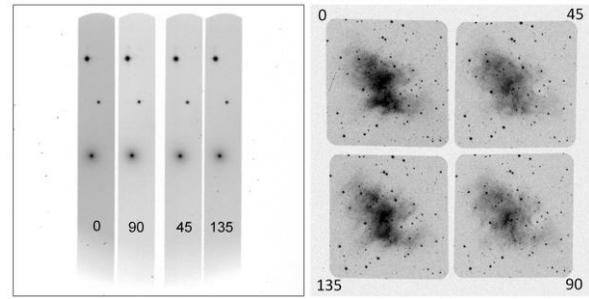

**FIGURE 3** Comparison of image polarimetry by StoP (*left*) and by MAGIC (*right*) captured to the same Andor CCD. 0, 45, 90, 135 — four directions of polarization (in degrees).

MAGIC as a focal reducer decreases the equivalent focal length, converting the telescope focal ratio of the optical system of the Zeiss-1000 telescope $F/13.3$ into the resulting $F/6.1$, which increases the field of view. Since the reduction of the telescope focal ratio is provided by a collimator and a camera, the optical scheme has a collimated (parallel) beam. It is possible to introduce into the beam a dispersing element in order to obtain low-resolution spectra (spectral resolution $R \sim 1000$) in the wavelength range of 400-740 nm with dispersion 0.2 nm/px, using a long slit with dimensions 2″ by 12′). The quadruple Wollaston prism makes it possible to use the entire useful field of view (6′.5 for each of the four directions of polarization). The photometric mode implies the 12′ FoV with a scale of 0″.45/px and the possibility of using various filters.

Fig. 3 shows for comparison the raw images in polarimetric mode with a wedged prism in StoP and a quadruple Wollaston in MAGIC. In both cases, the mask is introduced into the beam to avoid the overlapping of the images. However, the FoV given by masks differs sharply: in case of StoP, the rectangular mask FoV is only 0.9x5, while in MAGIC one obtain 6.5x6.5 images in the polarimetric mode. To apply the differential polarimetry technique (Afanasiev & Amirkhanyan,

2012) to significantly increase the accuracy of the measurement one need to observe the local standard stars in the same field with the investigated object. Therefore, in the case of small FoV of the rigidly fixed mask in StoP, a rotator is required for the target positioning. For MAGIC, a rotator is not so necessary; however, the number of the tasks for the long-slit observations can also require the adjusted orientation angle.

For low-level management, control of mechanical assemblies for switching filters, mirrors, moving the carriage, turning the device along a positional angle is carried out using a process control block (PCB) with a microprocessor developed by us. Both StoP and MAGIC contains inside a control PC-minicomputer with Windows operating system, connected to the Internet. For high-level management, control of exposures and configuration of devices in observations is carried out using special *StoP/MAGIC remote control* graphical interfaces, written by us using the IDL language.

## 3.2 Detector

The main light detector for both MAGIC and StoP is Andor iKon-L 936 CCD, BEX2-DD chip model[4]. A 5-stage peltier is used, which allows cooling using water to -100° C. In addition, air cooling is very convenient, and it cools down to -80°C. Deep depletion chip has practically no moire in the red part of the spectrum up to 1000 nm. This allows good images with the red filters and the absence of fringes throughout the all spectrum.

We examined the characteristics of the CCD in the laboratory. As an example, Fig. 4 shows the dependence of the dispersion value (or so-called Fano factor, see Afanasieva, 2016) on the flux given in electrons. This value shows the behavior of the CCD signal statistics in different regimes and gives an operating range of signal for optimal and correct

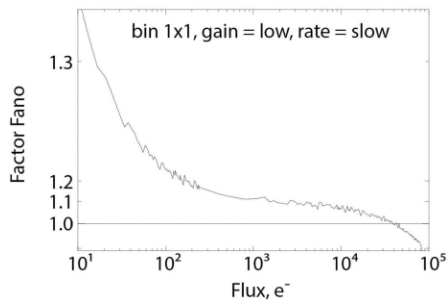

**FIGURE 4** Fano factor (dispersion index) for Andor CCD mode with low gain and slow rate setup.

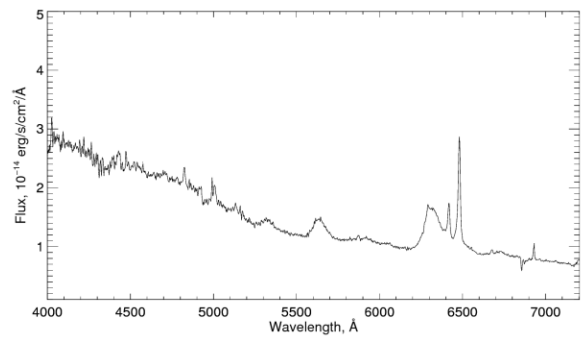

**FIGURE 6** First AGN spectrum using MAGIC: E1821+643.

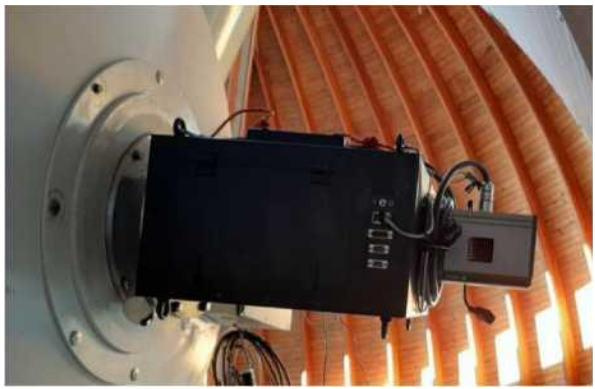

**FIGURE 5** Photo of MAGIC on the Zeiss-1000.

digital processing later on (see for more details Afanasieva, 2016).

Rigorous laboratory testing of the receiver is imperative to achieve correct acquisition levels and to understand optimal signal-to-noise ratio and statistics in observations.

### 3.3 Observations

The Stokes polarimeter as a prototype aboard the 1-m telescope obtained first light in the beginning of 2020. During the year, significant scientific results were obtained, which are described in the work by Afanasiev et al. (2021), as well as results on photometric reverberation mapping (Malygin, Shablovinskaya, Uklein, & Grokhovskaya, 2020). At the moment, StoP has been moved for observations with the SAO 60-cm telescope.

The first light of the MAGIC universal focal reducer was in September 2020 (see photo in Fig. 5). Observations are going on now with a gradual expansion of the scientific programs. The 1-meter telescope with the MAGIC instrument is used to solve monitoring polarimetric and photometric tasks, as well as monitoring the spectra of active nuclei in a limited mode.

Figure 6 shows an example of a spectrum obtained with MAGIC on 21 Sep 2020 with the 1-m telescope. Exposure time is equal 3600 sec, slit width is 2″. The operating range, taking into account the quantum efficiency curve of the CCD, is from 4000 to 7200A.

The strong point of MAGIC as a focal reducer and a polarimeter together is the possibility of polarimetry of extended sources with a field of view of 6.5 X 6.5. As an example, Figure 7 shows a composite image of M1 (Crab Nebula) and continuum polarimetry in the SED600 filter (with the central wavelength of about 600 nm and 25 nm bandwidth), where arrows indicate the direction and amplitude of polarization.

It is important to note, that the simultaneous observation of fluxes in four directions of polarization makes it possible to minimize the influence of the rapidly changing atmospheric parameters and to increase the accuracy of polarimetry in comparison with the usage of other analyzers.

For a starlike target up to 14 mag in medium-band filters with a seeing of 1″ for 20 minutes of total exposure, the photometry accuracy is better than 0.01 mag and the polarization accuracy is better than 0.6%. Observations are processed according to the algorithm described in the article by Afanasiev & Amirkhanyan (2012).

### 3.4 Perspectives

The MAGIC universal focal reducer has not yet been commissioned. For full-fledged operation, high-quality guiding with the 1-meter telescope is required. This is especially important for the long-slit spectroscopy. In addition, the current optical design has a strong beam separation and further research is required on this feature.

Currently, high-quality external guiding has not been implemented with the 1-meter telescope Zeiss-1000. Now our laboratory are preparing an automation of the rotator and the development of a polar guiding system inward. This project is in progress, therefore the observations are limited in terms of the exposures. Within non-ideal telescope tracking the maximum exposure for the direct images and polarimetry is 15 minutes, and about 10 minutes for spectroscopy.

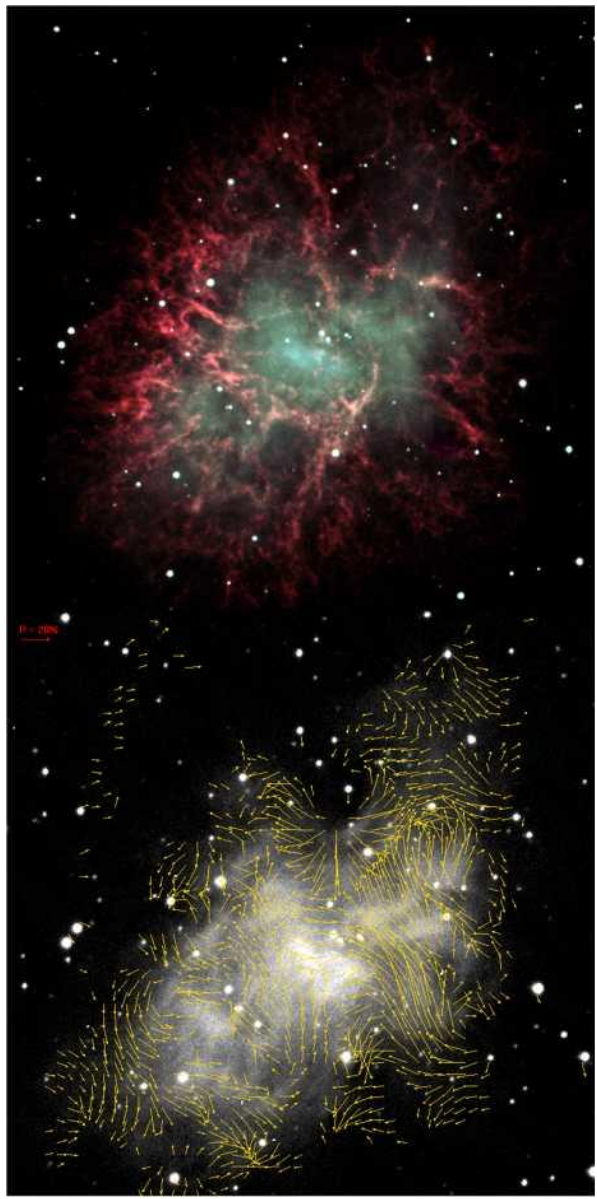

**FIGURE 7** M1 (Crab nebula) color *B+V*+Hα image *(top panel)* and continuum polarization map *(bottom panel)*.

In addition to guiding, a calibration module for illumination is also required to make continuous and arc calibration spectra in the long slit mode. Focusing of the collimator, as well as some flat fields for polarimetry, are performed using the dome flat on the telescope. In observations, flat fields are also made for photometry and polarimetry by the twilight sky.

One of the prospects for the development of MAGIC is the transition to automatic monitoring under the control of an observer, now in our project we are debugging the technique on the device, and the next creation of additional modules. An example of a universal MAGIC operating on the SAO 1-m telescope may be a good initial solution for observations with small telescopes.

## 4 | CONCLUSIONS

MAGIC is lightweight universal (multi-mode) focal 1:2.2 reducer that can be used on small telescopes to solve various observational tasks:

- Direct imaging in the Johnson-Cousins (UBVRI) photometric system and in the midband (SED) interference filters; the photometry in FoV ~ 12′ with a scale of 0″.45/pix (2 turrets, 9 positions each);

- Image polarimetry by quadrupole Wollaston prism with 6′.5 for each of the four directions of polarization;

- Long-slit 12′ X 2′ spectroscopy with a resolution of $R$ ~ 1000 in the 400-740 nm range.

In 2022 we plan to commission a rotator, a guide system and a calibration illumination module on the 1-meter SAO telescope for full-fledged operation MAGIC universal focal reducer.

## ACKNOWLEDGMENTS

This work was supported by the **Russian Scientific Foundation** (grant no. *20-12-00030* "Investigation of geometry and kinematics of ionized gas inactivegalacticnucleibypolarimetry methods").

Observations with the SAO RAS telescopes are supported by the **Ministry of Science and Higher Education of the Russian Federation** (including agreement No. *05.619.21.0016*, project ID RFMEFI61919X0016). The renovation of telescope equipment is currently provided within the national project "Science."

## Financial disclosure

None reported.

## Conflict of interest

The authors declare no potential conflict of interests.

# AUTHOR BIOGRAPHY

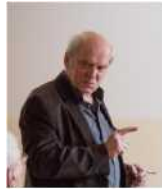

**Victor Afanasiev** (1947-2020)
Prof., main researcher of LSPEO, SAO RAS. The creator and developer of the astronomical instrumentation. The specialist in the field of active galaxies, 3D spectroscopy and polarimetry.

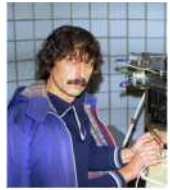

**Vladimir Amirkhanyan**
Senior researcher of LSPEO, SAO RAS, and Department of Radioastronomy, Moscow State University. The experienced engineer and astrophysicist, created the electro-mechanical schemes of all LSPEO instruments.

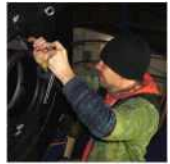

**Alexander Perepelitsyn**
Engineer of LSPEO, SAO RAS. Technical support and modernization of all LSPEO instruments.

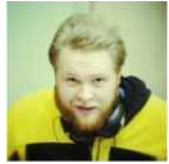

**Roman Uklein**
Researcher of LSPEO, SAO RAS. Works in various optical telescope studies of galaxies. Advanced in observational instruments and UI programming.

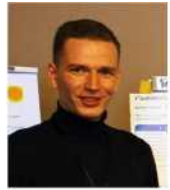

**Eugene Malygin**
PhD student of LSPEO, SAO RAS. Observer. Interested in studying SMBH in AGNs.

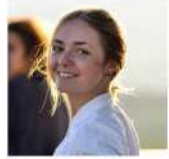

**Elena Shablovinskaya**
Junior researcher of LSPEO, SAO RAS. Experienced in optical polarimetry of central parts of AGNs.

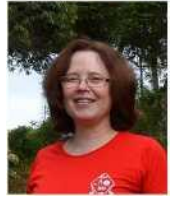

**Irina Afanasieva**
Head of the Advanced Design Laboratory, SAO RAS. Experienced in software development with a special focus on complex, reactive and event-driven applications for astronomical equipment.